\newcommand{\mymat}[1]{{\underline{\underline{\mathbf{#1}}}}}
\newcommand{\rv}{{\bf r}}
\newcommand{\rhat}{\hat{\rv}}
\newcommand{\rast}{{R^\ast}}
\newcommand{\nv}{{\bf n}}
\newcommand{\cv}{{\bf c}}
\newcommand{\qv}{{\bf q}}
\newcommand{\ev}{{\bf e}}
\newcommand{\nid}[2]{{\bf n}^{(#1)}\left(#2\right)}
\newcommand{\radavg}[1]{\left\langle #1 \right\rangle_r}
\newcommand{\defdir}{{\bf d}}

\newcommand{\rotmat}[1]{\mymat{D}^{(#1)}}
\newcommand{\eqref}[1]{Eq.\ref{#1}}

\documentstyle[aps,pre,twocolumn,graphicx]{revtex}
\begin{document}


\title{
Topological Defects in\\ Nematic Droplets of 
Hard Spherocylinders
}

\author{
Joachim Dzubiella, Matthias Schmidt, and Hartmut L{\"o}wen
}

\address{\small \hfill                                   \\
                Institut f{\"u}r Theoretische Physik II, \\ 
                Heinrich-Heine-Universit{\"a}t D{\"u}sseldorf,
                Universit{\"a}tsstra{\ss}e 1,            \\
                D-40225 D{\"u}sseldorf, Germany          \\
}

\date{14 July 2000, to be published in Phys. Rev. E}

\maketitle

\begin{abstract}
Using computer simulations we investigate the microscopic structure of
the singular director field within a nematic droplet.  As a
theoretical model for nematic liquid crystals we take hard
spherocylinders. To induce an overall topological charge, the
particles are either confined to a two-dimensional circular cavity
with homeotropic boundary or to the surface of a three-dimensional
sphere. Both systems exhibit half-integer topological point defects.
The isotropic defect core has a radius of the order of one particle
length and is surrounded by free-standing density oscillations. The
effective interaction between two defects is investigated.  All results
should be experimentally observable in thin sheets of colloidal liquid
crystals.
\end{abstract}



\narrowtext

\section{Introduction}

Liquid crystals (LC) show behavior intermediate between liquid and
solid. The coupling between orientational and positional degrees of
freedom leads to a large variety of mesophases. The microscopic origin
lies in anisotropic particle shapes and anisotropic interactions
between the particles that constitute the material.  The simplest,
most liquid-like  LC phase is the nematic phase where
the particles are aligned along a preferred direction while their
spatial positions are, like in an ordinary liquid, homogeneously
distributed in space. The preferred direction, called the nematic
director, can be macroscopically observed by illuminating a nematic
sample between crossed polarizers.

There are many different systems that possess a nematic phase.
Basically, one can distinguish between molecular LCs where the
constituents are molecules and colloidal LCs containing mesoscopic
particles, e.g., suspensions of tobacco mosaic viruses
\cite{vroege92}. Furthermore there is the possibility of
self-assembling rodlike micelles \cite{herbst93}, that can be studied
with small-angle neutron scattering \cite{froeba95}.

There are various theoretical approaches to deal with nematic liquid
crystals.  On a coarse-grained level one may use Ginzburg-Landau
theories, including phenomenological elastic constants. The central
idea is to minimize an appropriate Frank elastic energy with respect
to the nematic director field \cite{chandrasekhar77}.  Second, there
are spin models, like the Lebwohl-Lasher model, see, e.g., Refs.\ 
\cite{chiccoli90,berggren94bipolar,berggren94}. There the basic
degrees of freedom are rotators sitting on the sites of a lattice and
interacting with their neighbors.  The task is to sample appropriately
the configuration space.  The third class of models consists of
particles with orientational and positional degrees of freedom.
Usually, the interaction between particles is modelled by an anisotropic pair
potential.  Examples are Gay-Berne particles, e.g.\ 
\cite{allen96,stelzer95} and hard bodies, e.g., hard spherocylinders
(HSC) \cite{allen93}.  Beginning with the classical isotropic-nematic
phase transition for the limit of thin, long needles due to Onsager
\cite{onsager49}, our knowledge has grown enormously for the system of
hard spherocylinders.  The bulk properties have recently been
understood up to close packing.  The phase diagram has been calculated
by computer simulations \cite{bolhuis97frenkel}, density-functional
theory \cite{graf99} and cell theory \cite{MainzHSC}. There are
various stable crystal phases, like an elongated face-centered cubic
lattice with ABC stacking sequence, a plastic crystal, smectic-A
phase, nematic and isotropic fluid.  Besides bulk properties, one has
investigated various situations of external confinement, like nematics
confined to a cylindrical cavity \cite{bradac98} or between parallel
plates \cite{borstnik97,gruhn97}.  Also effects induced by a single
wall have been studied, like depletion-driven adsorption
\cite{sear98}, anchoring \cite{schuddeboom97}, wetting \cite{roij00},
and the influence of curvature\cite{groh99}.  Furthermore, solid
bodies immersed in nematic phases experience non-trivial forces
\cite{ramaswamy96,alouges99,poulin99}, and point defects experience an
interaction \cite{semenov99}.

Topological defects within ordered media are deviations from ideal
order, loosely speaking, that can be felt at an arbitrary large
separation distance from the defect position.  Complicated examples
are screw dislocations in crystalline lattices and inclusions in
smectic films \cite{pettey98}.  To deal with topological defects the
mathematical tools of homotopy theory may be employed \cite{mermin79}
to classify all possible structures. The basic ingredients are the
topology of both the embedding physical space and the order parameter
space.  For the case of nematics, there are two kinds of stable
topological defects in 3d, namely point defects and line defects,
whereas in 2d there are only point defects. These defects arise when
the system is quenched from the isotropic to the nematic state
\cite{hindmarsh95}. Also the dynamics have been investigated
\cite{wang98} experimentally.  On the theoretical side, there is the
important work within the framework of Landau theory by Schopohl and
Sluckin on the defect core structure of half-integer wedge
disclinations \cite{schopohl87} and on the hedgehog structure
\cite{schopohl88} in nematic and magnetic systems. The latter
predictions have been confirmed with computer simulations of lattice
spin models \cite{chiccoli95}.  The topological theory of defects has
been used to prove that a uniaxial nematic either melts or exhibits a
complex biaxial structure \cite{biscari97}.  Sonnet, Kilian and Hess
\cite{sonnet95} have considered droplet and capillary geometries using
an alignment tensor description.

The investigation of equilibrium topological defects in nematics has
received a boost through a striking possibility to stabilize defects
by imprisoning the nematic phase within a spherical droplet. The
droplet boundary induces a non-trivial effect on the global structure
within the droplet. Moreover, it can be experimentally controlled in a
variety of ways to yield different well-defined boundary conditions,
namely homeotropic or tangential ones.  One famous experimental system
are polymer-dispersed LCs. Concerning nematic droplets, there are
various studies using the Lebwohl-Lasher model
\cite{chiccoli90,berggren94bipolar,berggren94}.  There are
investigations of the droplet shape \cite{huang94,ambrozic97}, the
influence of an external field \cite{xu94}, and chiral nematic
droplets \cite{bajc95}, structure factor \cite{zapotocky99}, and ray
propagation \cite{reyes98}.  Also simulations of Gay-Berne droplets
have been performed \cite{emerson95}.  Other systems that exhibit
topological defects are nematic emulsions
\cite{poulin97science,stark99,lubensky98}, and defect gels in
cholesteric LCs \cite{zapotocky99science}. The formation of
disclination lines near a free nematic interface was reported
\cite{ignes-mullol99}.

In this work we are concerned with the microscopic structure of
topological defects in nematics.  We use a model for rod-like
particles with a pair-wise hard core interaction, namely hard
spherocylinders. It accounts for both, the orientational degrees of
freedom as well as the positional degrees of freedom of the particles
constituting the nematic.  Especially, it allows for mobility of the
defect positions.  This system is investigated with Monte Carlo
computer simulations.  There exist successful simulations of
topological line defects using hard particles, namely integer
\cite{andrienko00} and half-integer line defects \cite{hudson93}.

Here, we undertake a detailed study of the microscopic structure of
the defect cores focusing on the behavior of the local nematic order
and on the density field, an important quantity that has not been
studied in the literature yet.  As a theoretical prediction, we find
that the arising half-integer point defects are surrounded by an
oscillating density inhomogeneity.  This can be verified in
experiments.  We also investigate the statistical properties of two
defects interacting with each other extracting the distribution
functions of the positions of the defect cores and their orientations.
These are not accessible in mean-field calculations.  We emphasize
that both properties, the free-standing density wave which is due to
microscopic {\em correlations} and the defect position distribution
which is due to {\em fluctuations} cannot be accessed by a
coarse-grained mean-field type calculation.

The paper is organized as follows: In section \ref{SECmodel} our
theoretical model is defined, namely hard spherocylinders within a
planar spherical cavity and on the surface of a sphere. For
comparison, we also propose a simplified toy model of 
aligned rods. Section \ref{SECtools} is devoted to the analytical
tools employed, such as order parameter and density profiles.  Section
\ref{SECsimulation} gives details about the computer simulation
techniques used. The results of our investigation are given in section
\ref{SECresults} and we finish with concluding remarks and a
discussion of the experimental relevance of the present work in
section \ref{SECconclusions}.

\section{The Model}
\label{SECmodel}

\subsection{Hard Spherocylinders}
\begin{figure}
\begin{center}
   \includegraphics[width=0.77\columnwidth]{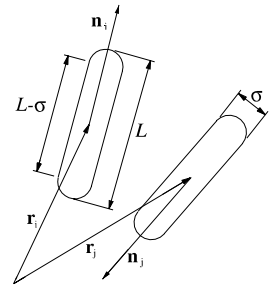}
\end{center}
\caption{
  Two hard spherocylinders with position coordinates $\rv_i$ and
  $\rv_j$, and orientations $\nv_i$ and $\nv_j$. The width
  of the particles is $\sigma$, the total rod length is denoted by
  $L$.  }
\label{FIGspherocylinder}
\end{figure}

We consider $N$ identical particles with center-of-mass position
coordinates $\rv_i=(r_{xi}, r_{yi})$ and orientations $\nv_i$,
where the index $i=1,\ldots,N$ labels the particles. Each particle has
a rod-like shape: It is composed of a cylinder of diameter $\sigma$
and length $L-\sigma$ and two hemispheres with the same diameter
capping the cylinder on its flat sides. In three dimensions (3d) this
geometric shape is called a spherocylinder, see
Fig.\ref{FIGspherocylinder} .  The 2d analog is sometimes called
discorectangle as it is made of a rectangle and two half discs. We
assume a hard core interaction between any two spherocylinders that
forbids particle overlap.  Formally, we may write
\begin{eqnarray}
  U(\rv_i, \nv_i; \rv_j, \nv_j) = \left \{
\begin{array}{ll}
    \infty &  {\rm if \; particles}\; i \; {\rm and} \; j \; {\rm overlap}\\
    0 &  {\rm else}
\end{array}
\right.
\end{eqnarray}
The geometric overlap criterion involves a sequence of elementary
algebraic tests. They are composed of scalar and vector products
between the distance vector of both particles and both orientation
vectors. The explicit form can be found e.g.\ in Ref.\ 
\cite{lowen94hsc}.  The bulk system is governed by two dimensionless
parameters, namely the packing fraction $\eta$, which is the ratio of
the space filled by the particle ``material'' and the system volume $V$.
In two dimensions it is given by $\eta=(N/V)(\sigma (L-\sigma)+\pi
\sigma^2/4)$.  The second parameter is the anisotropy $p=L/\sigma$
which sets the length-to-width ratio.  The bulk phase diagram in 3d
was recently mapped out by computer simulation \cite{bolhuis97frenkel}
and density-functional theory \cite{graf99}. The nematic phase is
found to be stable for anisotropies $p>5$. In 2d the phase diagram is
not known completely but there is an isotropic to nematic phase
transition for infinitely thin needles \cite{frenkel85}. The nematic
phase is also present in a system of hard ellipses
\cite{viellard-baron72,cuesta90} verified by computer simulations. In
2d the nematic-isotropic transition was investigated using
density-functional theory \cite{schoot97} and scaled-particle theory
\cite{schlacken98}. There is work about equations of state
\cite{maeso95}, and direct correlation functions \cite{chamoux98}
within a geometrical framework.

\subsection{Planar model}
\begin{figure}
\begin{center}
   \includegraphics[width=0.77\columnwidth]{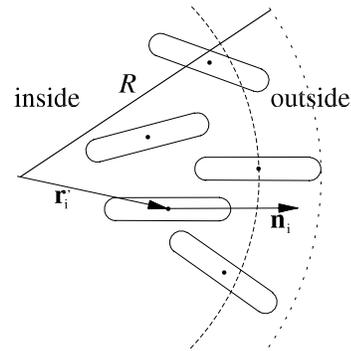}
\end{center}
\caption{
  Homeotropic boundary conditions for the planar droplet. The particle
  centers (points) are not allowed to cross a circle with diameter
  $R-L/2$ (dashed line). Then the shape of each particle lies inside a
  circle with radius $R$.  }
\label{FIGplanarBoundary}
\end{figure}

To align the particles near the system boundary homeotropically we
apply a suitably chosen external potential.  The particles are
confined within a spherical cavity representing the droplet shape. The
interaction of each HSC with the droplet boundary is such that the
center of mass of each particle is not allowed to leave the droplet,
see Fig.\ref{FIGplanarBoundary}.  The corresponding external potential
is given by
\begin{eqnarray}
  U_{\rm ext}(\rv_i) = \left \{
\begin{array}{lll}
   0 &  {\rm if} & |\rv_i| < R - L/2\\
  \infty &  {\rm else} &\\
\end{array}
\right.
\end{eqnarray}
where $R$ is the radius of the droplet and we chose the origin of the
coordinate system as the droplet center. The system volume is $V=\pi
R^2$. This boundary condition is found to induce a nematic order
perpendicular to the droplet boundary as the particles try to stick
one of their ends to the outside\cite{allen99homeotropic}. Hence the
topological charge is one. In the limit, $p=1$, we recover the
confined hard sphere system recently investigated in 2d
\cite{nemeth982d} and 3d \cite{nemeth983d,gonzalez97,gonzalez98}.

\subsection{Spherical model}
\begin{figure}
\begin{center}
    \includegraphics[width=0.77\columnwidth]{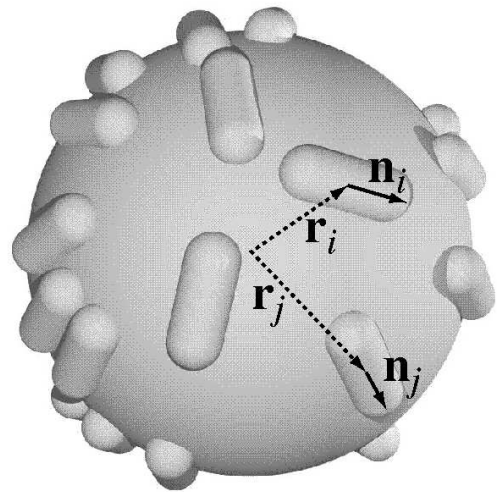}
\caption{
  Spherical system. Each particle with position $\rv_i$ and
  orientation $\nv_i$ is forced to lie tangentially on the surface
  of a sphere.  }
\label{FIGsphericalModel}
\end{center}
\end{figure}

A second possibility to induce an overall topological charge is to
confine the particles to a non-planar, curved space, which we chose to
be the surface of a sphere in three-dimensional space. The particles
are forced to lie tangentially on the sphere with radius $R$, see
Fig.\ref{FIGsphericalModel}.  Mathematically, this is expressed as
\begin{eqnarray}
| \rv_i | &= R, \\
\rv_i \cdot \nv_i &= 0.
\end{eqnarray}
The director field on the surface of a sphere has to have defects.
This is known as the ``impossibility of combing a hedgehog''. The
total topological charge \cite{mermin79} is two.  The topological
charge is a winding number that counts the number of times the
nematic director turns along a closed path around the defect.
It may have positive and negative, integer or half-integer values,
namely $0, \pm 1/2, \pm 1, \ldots$.

\subsection{Aligned Rods}
\begin{figure}
\begin{center}
    \includegraphics[width=0.77\columnwidth]{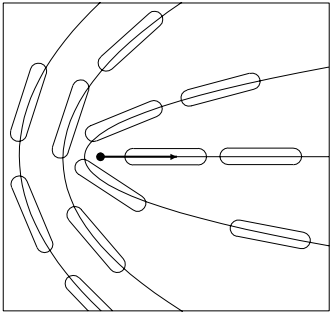}
\caption{
  Model of aligned rods. Each particle (discorectangles) has an
  orientation according to a prescribed director field (lines). The
  position of the arising $1/2$-defect is indicated by a filled
  circle, the orientation by an arrow.}
\label{FIGalignedRods}
\end{center}
\end{figure}

To investigate pure positional effects we study a further simplified
model where the orientation of each rod is uniquely determined by its
position. Therefore we consider an arbitrary unit vector field
$\nv(\rv)$ describing a given nematic order pattern.  In reality, the
particles fluctuate around this mean orientation. Here, however, we
neglect these fluctuation by imposing $\nv_i = \nv(\rv_i)$.  In
particular, we chose the director field to possess a singular defect
with topological charge $t$, see Fig.\ref{FIGalignedRods}. The
precise definition of this director field $\nid{t}{\rv}$ is postponed
to the next section (and given therein in \eqref{EQnid}.)  The case of
parallel aligned rods, $\nv={\rm const}$, has been used to study phase
transitions to higher ordered liquid crystals\cite{bohle96}.

\section{Analytical Tools}
\label{SECtools}
\subsection{Order parameters}
In order to analyze the fluctuating particle positions and
orientations, we probe against a  director field possessing
a topological defect with charge $t$. It is given by
\begin{equation}
  \nid{t}{\qv, \rv} = \rotmat{t}(\rv) \qv,
  \label{EQnid}
\end{equation}
where the rotation matrix is
\begin{eqnarray}
 \mymat{D}^{(t)}({\bf a}) &=&
  \left (
    \begin{array}{cc} 
      \cos \left( t \phi \right ) & 
      -\sin \left( t \phi \right )  \\
      \sin \left(t \phi \right ) & 
      \cos \left( t \phi \right )
    \end{array}
  \right),
\end{eqnarray}
with $\phi=\arctan(a_y/a_x)$, and ${\bf a}=(a_x,a_y)$ being a 2d
vector. The vector $\qv$ is the orientation of particles if one
approaches the defect along the $x$-direction.

As an order parameter, we probe the actual particle
orientations $\nv_i$ against the ideal ones
\begin{equation}
  S^{(t)}(\cv,\qv;r) = 2 \radavg{ \left[ \nv_i \cdot
  \nid{t}{\qv, \rv_i-\cv}  \right]^2 } - 1,\label{EQStcq}
\end{equation}
where the radial average is defined as $\radavg{\ldots} = \left
  \langle \sum_{i=1}^N \delta(|\rv'_i| - r) \ldots \right \rangle
\left/ \left \langle \sum_{i=1}^N \delta(|\rv'_i| - r) \right \rangle
\right.$, with $\rv'_i=\rv_i-\cv$ and $\langle \ldots \rangle$ is an
ensemble average. Normalization in \eqref{EQStcq} is such that usually
$0\leq S^{(t)}\leq 1$, where unity corresponds to ideal alignment, and
zero means complete dissimilarity with the defect of charge $t$ at
position $\cv$ and vector $\qv$, \eqref{EQnid}. (In general,
$-1\leq S^{(t)}<1$ is possible, where negative values indicate an
anti-correlation.)

If $\cv$ and $\qv$ are not dictated by general symmetry considerations
(e.g. $\cv=0$ because of the spherical droplet shape), we need to
determine both quantities. To that end we measure the similarity of an
actual particle configuration compared to a defect, \eqref{EQnid}. We
probe this inside a spherical region around $\cv$ with radius $\rast$
using
\begin{equation}
  I^{(t)}(\cv,\qv) = \frac{2}{(\rast)^2} \int_0^\rast dr \, r \, 
  S^{(t)}(\cv,\qv;r).
\end{equation}
where $\rast$ is a suitably chosen cutoff length. We maximize
$I^{(t)}(\cv,\qv)$ with respect to $\cv$ and $\qv$.  The value at the
maximium is
\begin{equation}
  \lambda^{(t)} = \max_{\cv,\qv} \{ I^{(t)}(\cv,\qv) \},
  \label{EQmax}
\end{equation}
and the argument at the maximum is $\qv^{(t)}$.

Before summarizing the quantities we compute during the simulation,
let us note that $\qv^{(t)}$ and $\lambda^{(t)}$ are eigenvector and
the corresponding (largest) eigenvalue of a suitable tensor. To see
this, we attribute each particle the general tensor
\begin{eqnarray}
  \mymat{Q_{\it i}}^{(t)}&=&
  2  \left ( 
    \mymat{D}^{(t)}  (\rv_i - \cv) \, \nv_i \otimes 
  \mymat{D}^{(t)} (\rv_i - \cv) \, \nv_i
  \right )-\mymat{1},\label{EQqi}
\end{eqnarray}
where $\otimes$ denotes the dyadic product, $\mymat{1}$ is the
identity matrix. Summing over particles gives 
\begin{equation}
\mymat{Q}^{(t)}= \sum_i
\mymat{Q_{\it i}}^{(t)}. 
\label{EQQsum}
\end{equation}
Note that for $t=0$ the usual bulk nematic
order parameter is recovered\footnote{The constants in
  Eq.\protect\ref{EQqi} depend on the dimensionality of the system
  and are different from 3d, where, e.g. $\mymat{Q}^{(0)}=(3/2)
  \sum_i \nv_i \otimes \nv_i - \mymat{1}/2$ holds.}.  The order
parameter profile, \eqref{EQStcq}, is then obtained as
\begin{equation}
S^{(t)}(\cv,\qv,r) = \radavg{\qv \cdot  \mymat{Q}^{(t)}
 \cdot \qv },
\end{equation}
and then the relation $\lambda^{(t)} \qv^{(t)} = \mymat{Q}^{(t)}
\qv^{(t)}$ holds, if the sum over $i$ in \eqref{EQQsum} is restricted
to particles located inside a spherical region of radius $\rast$
around $\cv$.

Let us next give three combinations of $t,\cv,\qv$ that apply to the
current model. First, we investigate the (bulk) nematic order, $t=0$.
We resolve this as a function of the distance from the droplet center,
hence $\cv=0$.  The nematic director $\qv^{(0)}$ is obtained from
\eqref{EQmax} with $\rast=R$ The order parameter, defined in
\eqref{EQStcq}, then simplifies to
\begin{equation}
  S^{(0)}(r) = 2 \radavg{\left( \nv_i \cdot \qv^{(0)} \right)^2} -1.
\end{equation}

Second, we probe for star-like order, hence $t=1$, $\cv=0$. As we do
not expect spiral arms of the star pattern to occur, we can set
$\qv=\ev_x$, where $\ev_x$ is the unit-vector in
$x$-direction. We can rewrite \eqref{EQStcq} as
\begin{equation}
  S^{(1)}(r) = 2 \radavg{\left( \nv_i \cdot \rhat_i \right)^2} -1,
\end{equation}
where $\rhat_i=\rv_i/|\rv_i|$.

Third, we investigate $t=1/2$ defects. To that end, we need to search
for $\cv$ and $\qv$, as these are not dictated by the symmetry of the
droplet. Hence we numerically solve \eqref{EQmax} with $\rast=2L$ (see
Sec.\ref{SECtechnicalissues}.) We obtain
\begin{equation}
  S^{(1/2)}(r) = 2 \radavg{\left( \nv_i \cdot 
      \nid{1/2}{\qv^{(1/2)}, {\rv_i - \cv^{(1/2)}}} \right)^2} -1.
\end{equation}

The distribution of the {\em positions} of the particles is analyzed
conveniently using the density profile $\rho(r)$ around $\cv$, which
we define as
\begin{eqnarray}
\rho(r)=\left \langle
  (2 \pi r)^{-1}
  \frac{1}{N}\sum_{i=1}^N
  \delta(| \rv_i-\cv | -r) \right \rangle.
\label{densityprofile}
\end{eqnarray} 
We consider two cases: The density profile around the center of the
droplet, i.e. $\cv=0$, and around the position of a half-integer
defect, $\cv=\cv_1, \cv_2$.

It is convenient to introduce a further direction of a $t=1/2$ defect by
\begin{eqnarray}
  \defdir= 
  \mymat{D}^{(\frac{1}{2})}\left(\qv^{(\frac{1}{2})}\right)
  \qv^{(\frac{1}{2})}.
\end{eqnarray}
The vector $\defdir$ is closely related to
$\qv^{(\frac{1}{2})}$ by a rotation operation, where the rotation
angle is the angle between $\qv^{(\frac{1}{2})}$ and the $x$-axis.
The direction $\defdir$ is where the field lines are radial; see the
arrow in Fig.\ref{FIGalignedRods}.

\subsection{Defect distributions}
For a given configuration of particles the planar nematic droplet has
a preferred direction given by the global nematic director
$\qv^{(0)}$.  Each of the two topological defects has a position
$\cv_i$ and an orientation $\defdir_i, i=1,2$.  These quantities can
be set in relation to each other to extract information about the
average defect behavior and its fluctuations. In particular, we
investigated the following probability distributions depending on a
single distance or angle.

Concerning single defect properties, we investigate the separation distance
from the droplet center,
\begin{eqnarray}
  P(r)=
  (2 \pi \, r)^{-1}
  \frac{1}{2} \sum_{i=1,2} 
  \left\langle
    \delta (|\cv_i|-r)
  \right \rangle,
\end{eqnarray}
and the orientation relative to the nematic director,
\begin{eqnarray}
  P(\theta)=
  \frac{1}{2} \sum_{i=1,2}
  \left \langle 
    \delta \left(
    \arccos \left (\defdir_i \cdot\qv^{(0)} 
    \right)-\theta
    \right)
  \right \rangle.
\end{eqnarray}

Between both defects there is a distance distribution,
\begin{eqnarray}
  P(c_{12})=
  (2 \pi \, c_{12})^{-1}
  \left\langle \delta (|\cv_{1}-\cv_{2}|-c_{12})  
  \right \rangle,
\end{eqnarray}
and an angular distribution between defect orientations,
\begin{eqnarray}
  P(\theta_{12})=
  \left \langle 
    \delta \left(
    \arccos \left(
      \defdir_1 \cdot \defdir_2
    \right)-\theta_{12}
    \right) 
  \right \rangle,
\end{eqnarray}
which can equivalently be defined with
$\qv^{(\frac{1}{2})}_1,\qv^{(\frac{1}{2})}_2$ by using the identity
$\arccos (\defdir_1 \cdot \defdir_2) = 2 \arccos
(\qv^{(\frac{1}{2})}_1 \cdot \qv^{(\frac{1}{2})}_2).$

\section{Computer Simulation}
\label{SECsimulation}
\subsection{Monte Carlo}
All our simulations were performed with the canonical Monte-Carlo
technique keeping particle-number N, volume V and temperature T
constant, for details we refer to Ref.\ \cite{allen_tildesley}.  To
simulate spherocylinders with only hard interactions, each Monte-Carlo
trial is exclusively accepted when there is no overlap of any
particles. One trial always consists of a small variation of position
and orientation of one HSC.

For the planar case the translation for the particle $i$ is
constructed by adding a small random displacement $\Delta\rv_i$ to
the vector $\rv_i$, similarly the rotation consists of adding a
small random vector $\Delta \nv_i $ to the direction $\nv_i$ with
$\Delta\nv_i \cdot \nv_i=0$.

To achieve an isotropic trial on the surface of the sphere, the
rotation matrix $\mymat{M}$ is applied simultaneously to the vectors
$\rv_i$ and $\nv_i$. It is defined as
\begin{eqnarray}
\mymat{M} := 
\left(
  \begin{array}{ccc} 
    1-c+\alpha^{2} c &
    \gamma s+\alpha\beta c& -\beta s+\alpha\gamma c\\ 
    -\gamma s+\beta\alpha c &1-c+\beta c & \alpha +\beta\gamma
    c \\ 
    \beta s +\gamma \alpha  c & -\alpha s+\gamma\beta  c &
    1-c+\gamma^{2} c 
  \end{array}
\right)
\end{eqnarray}

with $s=\sin \Delta\theta$ and ${c=1-\cos \Delta\theta}$.
$\alpha,\beta,\gamma$ are for every trial randomly chosen cartesian
coordinates of the unit vector specifying the rotation axis, ${\Delta
  \theta}$ is a small random angle.  With this method a simultaneous
translation and rotation is warranted by keeping the vectors $\rv_i$
and $\nv_i$ normalized and perpendicularly oriented.

The maximal variation in all cases is adjusted such that the
probability of accepting a move is about fifty percent.  The overlap
criteria were checked by comparing the second virial coefficient of
two- and three-dimensional HSC with simulation results, where the
excluded volume of two HSC were calculated.  Each of the runs
(I)-(VII) was performed with $5\cdot 10^{7}$ trials per particle.  One
tenth of each run was discarded for equilibration.  Especially the
strongly fluctuating distance distribution between both defects,
$P(c_{12})$, needs good statistics. All quantities were averaged over 25
partial runs, from which also error bars were calculated.

An overview of the simulated systems is given in
Tab.\ref{TABparameters}.  The systems (I)-(VII) are planar. System (I)
is the reference. To study finite-size effects, system (II) has half
as many particles, and system (III) has twice as many particles as
(I). To investigate the dependence on the thermodynamic parameters,
system (IV) has a lower packing fraction $\eta$, and system (V) has a
higher one compared to system (I). The other thermodynamic parameter
is the anisotropy, which is smaller for system (VI) and higher for
system (VII) compared to the system (I). To keep the nematic phase
stable for the short rods of system (VI), the packing fraction $\eta$
had to be increased.  The packing fraction of the dense system (V) is
$\eta=0.4143$.  The spherical system has the same number of particles
$N$, packing fraction $\eta$ and anisotropy $p$ as the reference (I).
The radius of the sphere is half the radius of the planar droplet. The
aligned rod model has the same parameters as the reference system (I).

\subsection{Technical issues}
\label{SECtechnicalissues}
We discuss briefly a projection method for the spherical problem and a
search algorithm to find defect positions.

In order to perform calculations for the spherical system all
interesting vectors in three dimensions are projected to a
two-dimensional plane.  Imagine a given vector $\cv$ from the
middle of the sphere pointing to an arbitrary point of the surface. We
convert a position $\rv_i$ and orientation $\nv_i$ to the
vectors $\rv_i^{\rm p}$ and $\nv_i^{\rm p}$ in a plane
perpendicular to $\cv$ through
\begin{eqnarray}
\rv_i^{\rm p}=
\rv_i-(\cv  \cdot \rv_i)\cv,
\\
\nv_i^{\rm p}=
\nv_i-(\cv  \cdot \nv_i)\cv.
\end{eqnarray}
After obtaining a set $\{\rv^{\rm p}_i,\nv^{\rm p}_i\}$
of three dimensional vectors on this way, we transform them into a set
of two dimensional vectors by typical algebraic methods. As reference
the projection of the $\mathbf x$ unit vector of the fixed three
dimensional coordinate system is always the x-orientation of the ``new''
coordinate-system in two dimensions.  The results show that curvature
effects are small.

To investigate the radial structure and interactions of the
disclinations it is necessary to localize the centers of the two point
defects.  As described in the last section, the
$\lambda^{(\frac{1}{2})}$-parameter measures the degree of order of a
half-integer defect in a chosen area, so the task is to find the two
maxima of $\lambda^{(\frac{1}{2})}$ in the droplet.  In the planar
case, we do this search with the following algorithm: A circular
test-probe samples the droplet on a grid with a grid spacing of $5
\sigma$. At this points all the particles in the circle are taken to
calculate $\lambda^{(\frac{1}{2})}$ in the described way.  After
sampling the grid both maxima are stored and for every maximum a
refining Monte-Carlo-search is performed. The surrounding of size of
the grid spacing is randomly sampled and the probe is only moved when
$\lambda^{(\frac{1}{2})}$ increases. The search is stopped when the
probe does not move for 200 trials.  In the spherical case the method
is the same, but the grid is projected onto the sphere surface and the
calculations of $\lambda^{(\frac{1}{2})}$ were performed with
projected two-dimensional vectors as described before.

It is important to chose an adequate radius $\rast$ for the probe. If
$\rast$ is too large, the probe overlaps both defects. As they have
opposite orientations on the average, the located point of the maximum
deviates from the point we are interested in.  If the $\rast$ is too
small, an ill-defined position results, as fluctuations become more
important.  The simulation results show that a good choice is
$\rast=2L$.  Although this definition contains some freedom, we find
the defect position to be a robust quantity. A detailed discussion is
given in the following section.

\section{Results}
\label{SECresults}
\subsection{Order within the droplet}
\begin{figure}
\begin{center}
    \includegraphics[width=0.77\columnwidth]{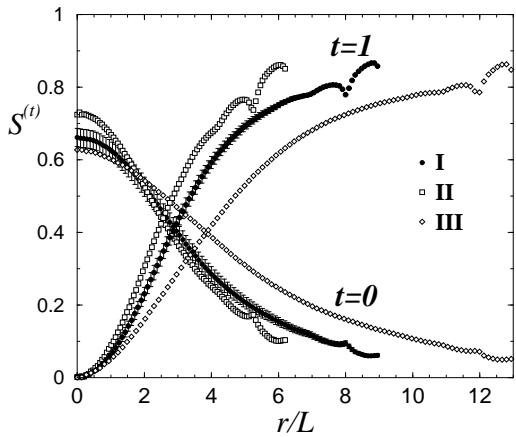}
\caption{
  Nematic order parameters $S^{(t)}$ as a function of the radial
  distance $r$ from the droplet center, scaled by the rod length $L$.
  Star order $S^{(1)}$ and bulk order $S^{(0)}$ is shown. System (I)
  is reference, (II) has halved and (III) has doubled particle number.
  See Tab.\ref{TABparameters} for a compilation of system
  parameters. Error bars are only given for (I).}
\label{FIGscissor} 
\end{center}
\end{figure}

Let us discuss the order parameters $S^{(t)}$ as a function of the radial
distance from the center of the droplet; see Fig.\ref{FIGscissor}.
$S^{(0)}$ is the usual bulk nematic order parameter, but radially
resolved. It reaches values of 0.6-0.75 in the middle of the droplet,
$r < 2 L$, indicating a nematic portion that breaks the global
rotational symmetry of the system. For $r>3 L$, $S^{(0)}$ decays to
values slightly larger than the isotropic value of $0$. The decrease,
however, is not due to a microscopically isotropic fluid state, as can
be seen from the behavior of $S^{(1)}$.  This quantity indicates
globally star-like alignment of particles for $r>3L$. It vanishes in
the nematic ``street'' in the center of the droplet. The distance
where $S^{(0)}$ and $S^{(1)}$ intersect is an estimate for the defect
positions.  In Fig.\ref{FIGscissor}, the finite size behavior of
$S^{(t)}$ is plotted for particle numbers $N=1004,2008,4016$
corresponding to systems (II), (I), (III). There is a systematic shift
of the intersection point of $S^{(0)}$ and $S^{(1)}$ to larger values
as the system grows, the numerical values are $r/L=2.54,2.91,3.87$.
However, if $r$ is scaled by the droplet radius $R$, a slight shift to
smaller values is observed as the system size grows.  Keeping the
medium-sized system (I) as a reference, we have investigated the
impact of changing the thermodynamic variables.  For different packing
fractions, $\eta$=0.2894 (IV), 0.3321 (I), 0.4143 (V), we found that
the intersection distances are $r/L$=3.90, 2.91, 1.43.  In the bulk,
upon increasing the density the nematic order grows.  Here, this
happens for the star-order $S^{(1)}$. But this increase happens on the
cost of the nematic street (see $S^{(0)}$) at small $r$-values.
Increasing $\eta$ leads to a compression of the inhomogeneous,
interesting region in the center of the droplet.  A similar effect can
be observed upon changing the other thermodynamic variable, namely the
anisotropy $p$.  The nematic street is compressed for longer rods,
$p=31$(VII), $r/L$=1.33.  Shorter rods, $p=16$, need a higher density
to form a nematic phase, so the values for systems (I), $r/L$=3.16,
and (VI), $r/L$=2.91, are similar, as both effects cancel out.

The behavior of $S^{(1)}$ is similar to the findings for a
three-dimensional droplet, where a quadratic behavior near $r=0$ was
predicted within Landau theory \cite{schopohl88}. A simulation study
using the Lebwohl-Lasher model \cite{chiccoli95} confirmed this
finding and revealed that a ring-like structure that breaks the
spherical symmetry is present.  A comparison to the results for a 3d
capillary by Andrienko and Allen \cite{andrienko00} seems
qualitatively possible as they find alignment of particles
predominantly normal to the cylinder axis. Their findings are
consistent with the behavior of $S^{(1)}$.  Although our system is
simpler as it only has two spatial dimensions, we could also establish
the existence of a director field that breaks the spherical symmetry
by considering the order parameter $S^{(0)}$.

\begin{figure}
\begin{center}
    \includegraphics[width=0.77\columnwidth]{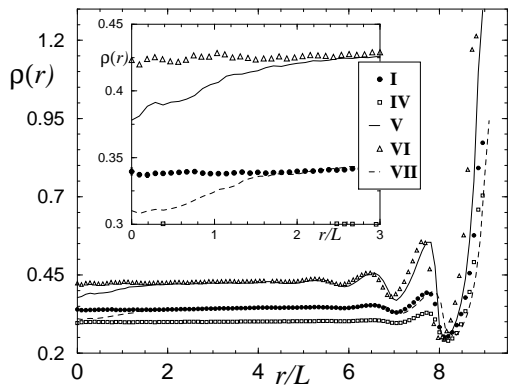}
\caption{
  Radially resolved density profiles $\rho(r)$ as a function of the
  distance from the droplet center $r$ scaled by the particle length
  $L$.  System (I) is reference, compared to lower (IV) and higher (V)
  packing fraction, and lower (VI) and higher (VII) anisotropies.
  The inset shows the behavior near the origin where a density
  decrease for (V) and (VII) appears for $r<2L$.  }
\label{FIGrhoglobal}
\end{center}
\end{figure}

Having demonstrated that the system exhibits a broken rotational
symmetry, we have to assure that no freezing into a smectic or even
crystalline state occurs. Therefore we plot radial density profiles
$\rho(r)$, where $r$ is the distance from the droplet center, in
Fig.\ref{FIGrhoglobal}.  The density shows pronounced oscillations for
large $r$ near the boundary of the system. They become damped upon
increasing the separation distance from the droplet boundary and
practically vanish after two rod lengths for intermediate density and
four rod lengths for high density. Approaching the droplet center,
$r=0$, the density reaches a constant value for the weakly nematic
systems (I), (IV), and (V).  For the strongly nematic systems, (V)
with high density and (VII) with large anisotropy, a density decay at
the center of the droplet occurs. This effect is not directly caused
by the boundary as the density oscillations due to packing effects are
damped.  It is merely due to the topological defects present in the
system. Quantitatively, the relative decrease is $\left[\rho(3
L)-\rho(0)\right]/\rho(3 L)$ = 0.11 (V), 0.09 (VII). The finite-size
corrections for systems (II) and (III) are negligible.

From both, the scissor-like behavior of the nematic order
(Fig.\ref{FIGscissor}) and from the homogeneity of the density profile
away from the system wall (Fig.\ref{FIGrhoglobal}), we conclude that
the system is in a thermodynamically stable nematic phase, and seems
to contain two  topological defects with charge~1/2.

In a 2d bulk phase, two half-integer (1/2) defects are more stable
than a single integer (1) defect, as the free energy is proportional
to the square of the charge.  However, in the finite system of the
computer simulation that is also affected by influence from the
boundaries, it could also be possible that the defect pair
merge into a single one \cite{andrienko00,sonnet95}.

\begin{figure}
\begin{center}
    \includegraphics[width=0.77\columnwidth]{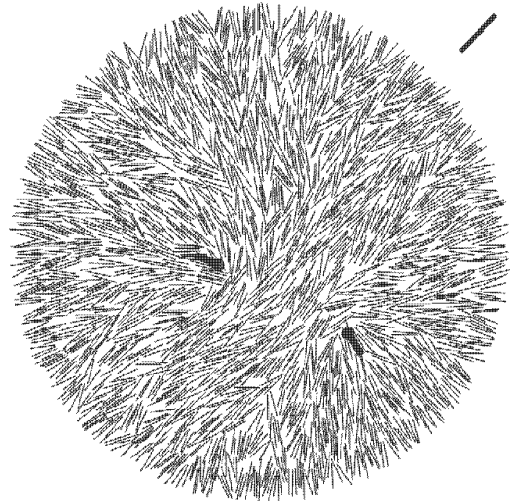}
\caption{
  Snapshot of a typical particle configuration for the planar system (I).
  The particles are rendered dark.  The two black symbols inside the
  droplet indicate positions and orientations of defects. The black
  bar outside the droplet indicates the global nematic director
  $\qv^{(0)}$.  }
\label{FIGsnapshotsPlanar}
\end{center}
\end{figure}

\begin{figure}
\begin{center}
  \includegraphics[width=0.77\columnwidth]{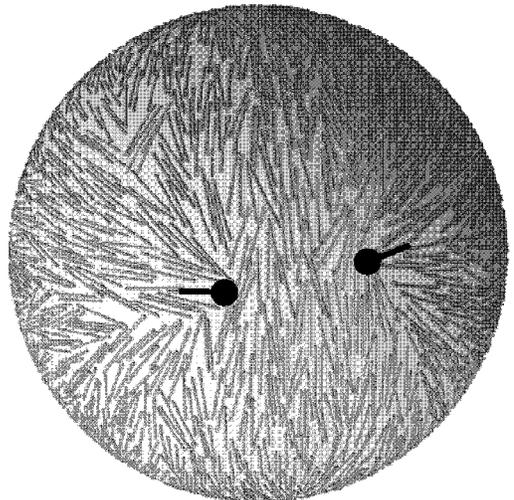}
\caption{
  Snapshot of a typical particle configuration for the spherical
  system.  The particles are rendered dark. There is one 1/2-defect
  on the left side and one on the right side. They point away from
  each other.  }
\label{FIGsnapshotsSpherical}
\end{center}
\end{figure}

Next we investigate the defect positions and their orientations. To
illustrate both, a snapshot of a configuration of the planar system is
shown in Fig.\ref{FIGsnapshotsPlanar} (I).  One can see the coupling
of the nematic order from the first layer of particles near the wall
to the inside of the droplet.  The particles near the center of the
droplet are aligned along a nematic director (indicated by the bar
outside the droplet).  The two emerging defects are depicted by
symbols.  See Fig.\ref{FIGsnapshotsSpherical} for a snapshot of the
spherical system.  There the total topological charge is not induced
by a system boundary but by the topology of the sphere itself.

\subsection{Defect core}
\begin{figure}
\begin{center}
    \includegraphics[width=0.77\columnwidth]{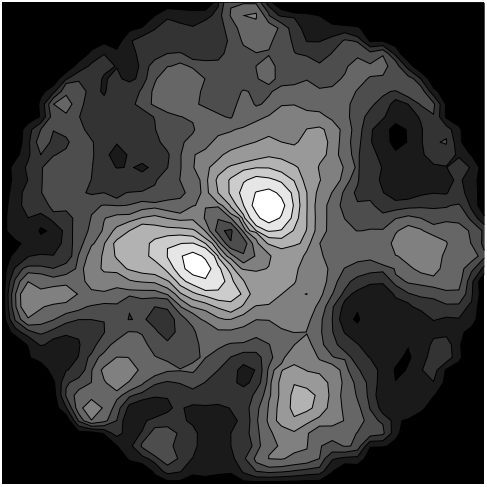}
\caption{
  Order parameter $\lambda^{(\frac{1}{2})}$ as a function of
  spatial coordinates $r_x, r_y$. Bright areas correspond to large
  values, dark areas correspond to small values of $\lambda^{(\frac{1}{2})}$.
  The two bright spots near the center are identified as topological
  defects, the gray islands as bulk defects.}
\label{FIGcontourplot}
\end{center}
\end{figure}

The positions of the defects are defined by maxima of the
$\lambda^{(\frac{1}{2})}$ order parameter, see Section \ref{SECtools}
for its definition. In Fig.\ref{FIGcontourplot},
$\lambda^{(\frac{1}{2})}$ is plotted as a function of the spatial
coordinates $r_x$ and $r_y$ for one given configuration.  There are
two pronounced maxima, indicated by bright areas, which are identified
as the positions of the defect cores $\cv_1$ and $\cv_2$.
There are several more local maxima appearing as gray islands.  These
are identified as statistical fluctuations already present in the bulk
nematic phase.

\begin{figure}
\begin{center}
    \includegraphics[width=0.77\columnwidth]{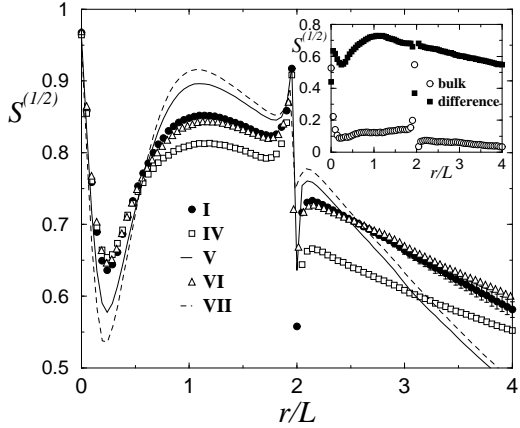}
\caption{
  Order parameter profiles $S^{(\frac{1}{2})}$ around the defect center as a
  function of the scaled distance $r/L$ from the defect center.  The
  reference system (I) is to be compared with lower (IV) and higher
  (V) packing fraction, and lower (VI) and higher (VII) anisotropies.
  The inset shows $S^{(\frac{1}{2})}$ for bulk defects and for the difference 
  between (I) and the bulk.
  }
\label{FIGhump}
\end{center}
\end{figure}

A drift of the positions of a defect core was also reported in
\cite{chiccoli95}. Here we follow this motion, to investigate the
surrounding of the defects. The order parameter $S^{(\frac{1}{2})}$ is
radially resolved around the defect position in Fig.\ref{FIGhump}. It
has a pronounced maximum around $r = 1.2 L$.  For smaller distances it
decreases rapidly due to disorder in the core region. For larger
distances the influence from the second defect partner decreases the
half-integer order $S^{(\frac{1}{2})}$. Increasing the overall
density, and increasing the anisotropy leads to a more pronounced
hump. The finite-size corrections, (II), (III), and the boundary
effects (sphere) are negligible. However, the curves show two
artifacts: A rise near $r=0$ and a jump at the boundary of the search
probe, $r=2L$.  In the inset the profile around a bulk defect is
shown. It has a plateau value inside the probe, $r<2L$, and vanishes
outside.  If we subtract this contribution from the pure data (I),
continuous behavior at $r=2L$ can be enforced.

However, the model does not account for 3d effects like the ``biaxial
escape'', namely the sequence planar uniaxial - biaxial - uniaxial
with increasing distance from the core center \cite{sonnet95}, as the
particles are only 2d rotators.  Schopohl and Sluckin
\cite{schopohl87} found an interface-like behavior between the inner
and outer parts of a disclination line in 3d.  In our system, we do
not find a sign of an interface between the isotropic core and the
surrounding nematic phase. This might be due to a small interface
tension and a very weak bulk nematic-isotropic phase transition.

\begin{figure}
\begin{center}
    \includegraphics[width=0.77\columnwidth]{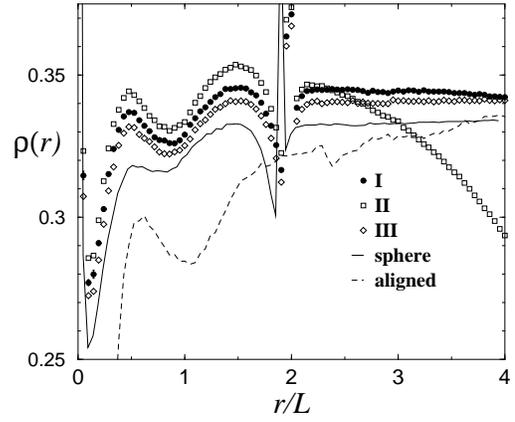}
\caption{
  Density profile as a function of the distance from the defect
  center.  System (I) is reference, (II) has fewer particles, (III)
  has more. The spherical and aligned models are shown.}
\label{FIGrhodefect1}
\end{center}
\end{figure}

\begin{figure}
\begin{center}
    \includegraphics[width=0.77\columnwidth]{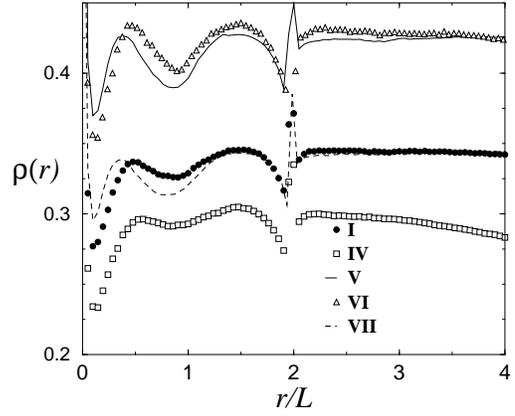}
\caption{
  Same as Fig.\ref{FIGrhodefect1}, but for lower (IV) and higher (V)
  packing fraction and shorter (VI) and longer particles (VII),
  compared to system (I).}
\label{FIGrhodefect2}
\end{center}
\end{figure}

By radially resolving the probability of finding a particle around a
defect center, we end up with density profiles depicted in
Fig.\ref{FIGrhodefect1}.  The defect is surrounded by density
oscillations with a wavelength of the particle length. The finite-size
dependence is small. To estimate the influence from the system wall,
one may compare with the spherical system.  It shows slightly weaker
oscillations. This might be due to curvature effects, as the effective
packing fraction is slightly smaller as the linear particles may
escape the spherical system.  The toy model of aligned rods also
exhibits a non-trivial density profile, showing a decrease towards
small distance and oscillations compared to rotating rods.  In all
cases the first peak has a separation distance of half a particle
length from the defect center. The second peak appears at $r=3/2L$.
Again the search probe induces an artificial structure near $r=2L$.
From this analysis, we can conclude that the oscillations are due to
packing effects.  The density oscillations become more pronounced at
higher density, and for larger anisotropy, see
Fig.\ref{FIGrhodefect2}.

\subsection{Defect position}
\begin{figure}
\begin{center}
    \includegraphics[width=0.77\columnwidth]{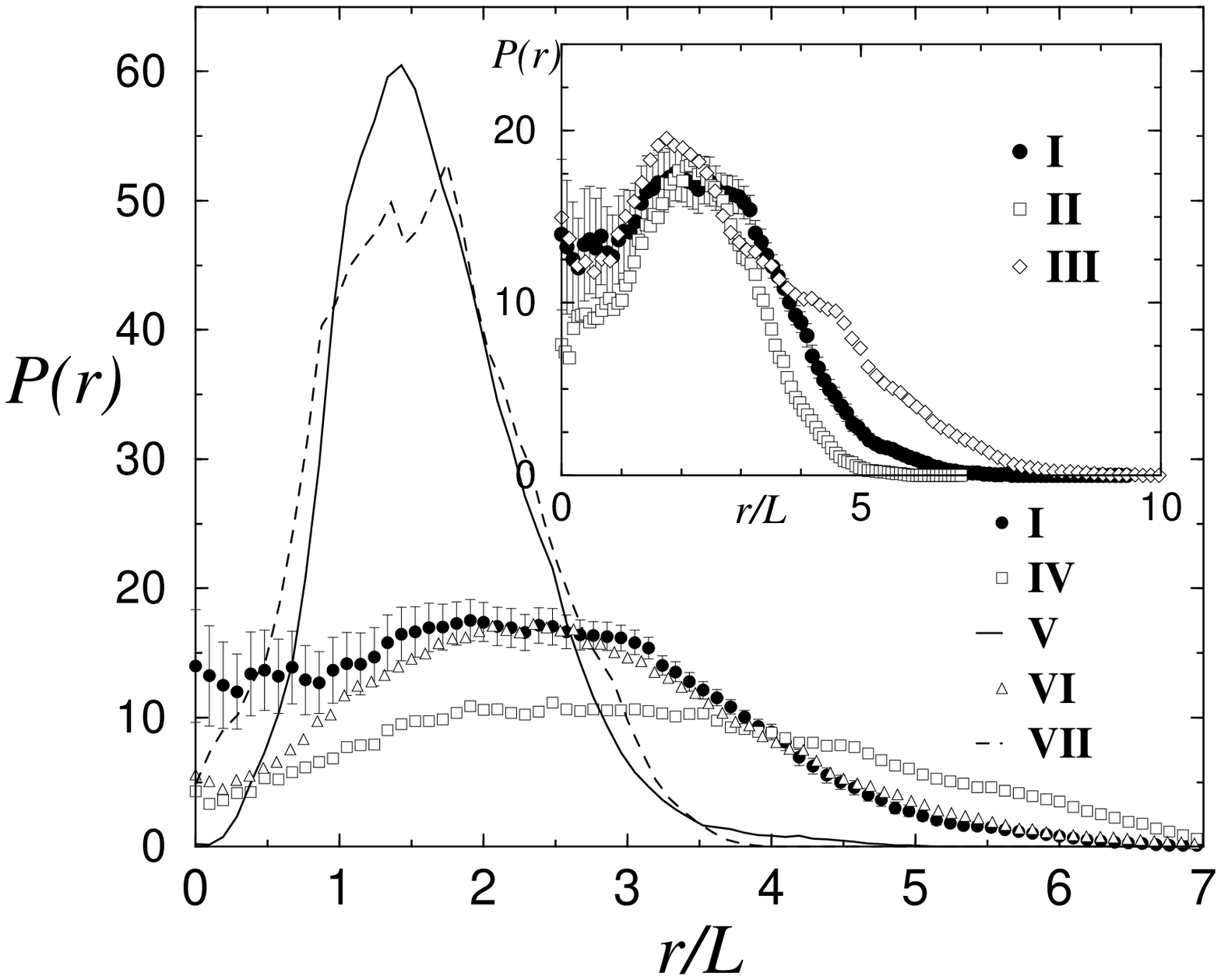}
\caption{
  Probability distribution $P(r)$ for the distance of a defect from
  the center of the droplet $r/L$, for lower (IV) and higher (V)
  packing fraction and shorter (VI) and longer particles (VII),
  compared to system (I). The inset shows the the finite size behavior
  for halved (II) and doubled (III) particle number. }
\label{FIGonedefDistance}
\end{center}
\end{figure}

In the planar system, each defect is characterized by its radial
distance $r$ from the center, and the angle $\theta$ between its
orientation and the global nematic director $\qv^{(0)}$.  We discuss
the probability distributions of these quantities.  In
Fig.\ref{FIGonedefDistance} the distribution for finding the defect at
a distance $r$ from the center is shown.  Generally, the distributions
are very broad. This indicates {\em large mobility} of the defects.
Changing the thermodynamical variables has a large effect. For the
stronger nematic systems (V) and (VII), the distribution becomes
sharper with a pronounced maximum at $r=1.5L$. Decreasing the
anisotropy weakens the nematic phase, so system (IV) has a very broad
distribution.  The inset shows that the distribution becomes broader
upon increasing system size.

\subsection{Interactions between two defects}
A complete probability distribution of both positions of the defect
cores can be regarded as arising from an effective interaction
potential $V_{\rm eff}(\cv_1, \cv_2)$ between the defects. The latter
play the role of quasi-particles. The effective interaction arises
from averaging over the particle positions while keeping the defect
positions constant.  The effective interaction and the probability
distribution are related via $P({\cv_1},{\cv_2}) \propto \exp (- \beta
V_{\rm eff}(\cv_1, \cv_2) )$.

\begin{figure}
\begin{center}
    \includegraphics[width=0.77\columnwidth]{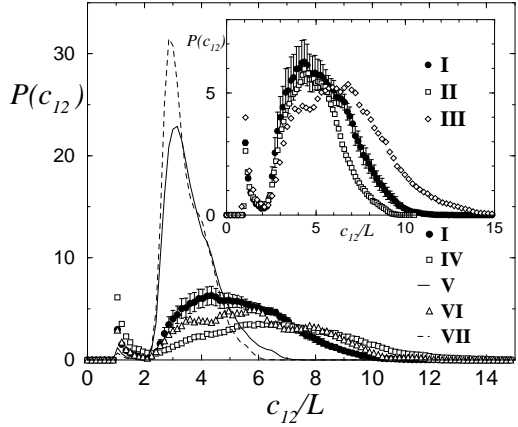}
\caption{
  Probability distribution $P(c_{12})$ for the separation distance
  between both defect positions scaled by the particle length for
  lower (IV) and higher (V) packing fraction and shorter (VI) and
  longer spherocylinders (VII), as compared to system (I). The inset
  shows the finite size behavior for halved (II) and doubled (III)
  particle number compared to (I).}
\label{FIGtwodefDistance}
\end{center}
\end{figure}

Instead of the full probability distribution, we show its dependence
on the separation distance between both defects and on their relative
orientation. In Fig.\ref{FIGtwodefDistance} the probability
distribution of finding two defects at a distance $c_{12}$ is shown.
It has small values for small as well as large  $c_{12}$.
Hence at small distances the defects repel each other.
At large distances their effective interaction is attractive.
Increasing the nematic order by increasing the density (V) or rod
length (VII) causes the average defect separation distance to shrink.
The rise near $r/L=1$ is an artifact: These are events where the
search algorithm does not find two different defects, but merely finds
the same defect two times. To avoid the problem a cutoff at $r=L$ was
introduced. The finite size behavior is strong; see the inset. The
large system (III) allows the defects to move further away from each
other, whereas in the smaller system (II) they are forced to be closer
together.  However, from the simulation data, it is hard to obtain the
behavior in the limit $R/L \to \infty$.

This is somewhat in contrast to the phase diagram of a 3d capillary
\cite{sonnet95} containing isotropic, planar-radial and
planar-polar structures, if one is willing to identify the dependence
on temperature with our athermal system.  There it was found that
the transition from the planar-polar to the planar-radial structure
happens upon increasing the temperature (and hence decreasing the
nematic order).

\begin{figure}
\begin{center}
    \includegraphics[width=0.77\columnwidth]{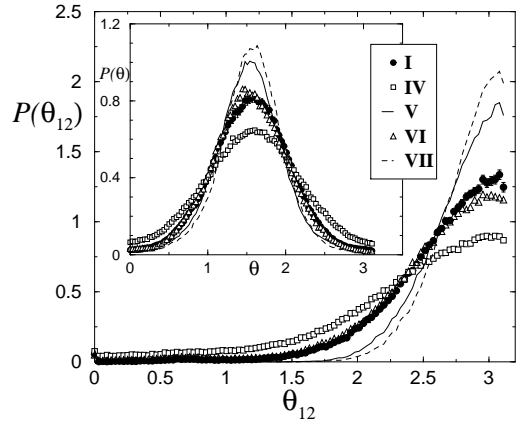}
\caption{
  Probability distribution $P(\theta_{12})$ for the difference angle
  between both defect orientations. The reference system (I) is to be
  compared with lower (IV) and higher (V) packing fraction, and lower
  (VI) and higher (VII) anisotropies. The inset shows the distribution
  $P(\theta)$ of the difference angle between the direction of one of the
  defects and the global nematic director for the same parameters.}
\label{FIGdefAngle}
\end{center}
\end{figure}

The difference angle $\theta_{12}$ between both defect orientations in
the planar system, see Fig.\ref{FIGdefAngle}, is most likely $\pi$,
hence the defects point on average away from each other. However, the
orientations are not very rigid.  For the least ordered system (IV)
there is still a finite probability of finding the defects with a
relative orientation of 90 degrees! Even for the strongly nematic
systems (V) and (VII) the angular fluctuations are quite large.  The
inset in Fig.\ref{FIGdefAngle} shows the distribution of the angle
$\theta$ between the defect orientation and the global nematic
director.  A clear maximum near $\pi/2$ occurs. Again, the
distributions become sharper as density or anisotropy increase.

\subsection{Outlook}
\begin{figure}
\begin{center}
    \includegraphics[width=0.77\columnwidth]{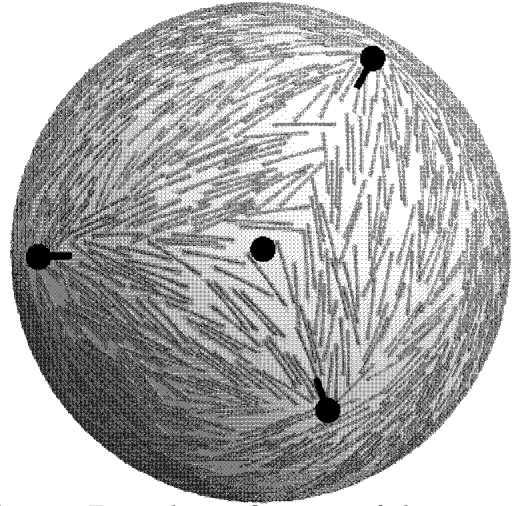}
\caption{
  Triangular configuration of three positive defects around a
  spontaneously formed negatively charged defect (central dot).}
\label{FIGsnapshotsOutlook3}
\end{center}
\end{figure}

Finally, it is worth mentioning that the spherical system still
contains surprises.  See Fig.\ref{FIGsnapshotsOutlook3} for an
unexpected configuration, namely an assembly of three positive
1/2-defects sitting at the corners of a triangle and a negative
-1/2-defect in its center. This is remarkable, because the negative
defect could annihilate with one of the outer positive defects.

In all cases, integer defects seem to dissociate into half-integer
defects.  The complete equilibrium defect distribution of hard
spherocylinders lying tangentially on a sphere remains an open
question.

\section{Conclusions}
\label{SECconclusions}
In conclusion, we have investigated the microscopic structure of
topological defects of nematics in a spherical droplet with
appropriate homeotropic boundary and for particles lying on the
surface of a sphere. We have used hard spherocylinders as a model
system for a lyotropic nematic liquid crystal. This system allows us
to study the statistical behavior of the microscopic rotational and
positional degrees of freedom. For this system we find half-integer
topological point defects in two dimensions to be stable. The defect
core has a radius of the order of one particle length. As an important
observation, the defect generates a free-standing density oscillation.
It possesses a wavelength of one particle length.  Considering the
defects as fluctuating quasi-particles we have presented results for
their effective interaction.

The microscopic structure revealed by radially resolving density and
order parameter profiles around the defect position is identical for
the planar and the spherical system. 

An experimental investigation using anisotropic colloidal particles
\cite{zahn94,zahn99} like tobacco mosaic viruses or carbon nanotubes
is highly desirable to test our theoretical predictions. Then larger
accessible system sizes can be exploited. Also of interest is the
long-time dynamical behavior of the motion of topological defects. The
advantage of colloidal systems over molecular liquid crystals is the
larger length scale that enables real-space techniques like digital
video-microscopy to be used.

From a more theoretical point of view it would be interesting to
describe the microstructure of topological defects within the
framework of density functional theory.  Using phenomenological
Ginzburg-Landau models, one could take the elastic constants of the
HSC model as an input, and could calculate the defect positions and
check against our simulations.

Finally we note that we currently investigate the three-dimensional
droplets that are filled with spherocylinders. In this case more
involved questions appear, as both, point and line defects, may
appear.

\vspace{1cm}

{\bf Acknowledgment.} It is a pleasure to thank J{\"u}rgen Kalus,
Karin Jacobs, Holger Stark, and Zsolt N{\'e}meth for useful
discussions, and Holger M. Harreis for a critical reading of the
manuscript.


\begin{table}
\begin{center}
\begin{tabular}{|c|c|c|c|r|}\hline 
System & $N$ & $p$ & $\eta$  & $2R/L$\\ \hline \hline
I   & 2008 & 21 & 0.3321 & $19.05$ \\
II  & 1004 & 21 & 0.3321 & $13.41$ \\
III & 4016 & 21 & 0.3321 & $26.94$ \\
IV  & 1750 & 21 & 0.2894 & $19.05$ \\
V   & 2500 & 21 & 0.4143 & $19.05$ \\
VI  & 1855 & 16 & 0.4143 & $18.75$ \\
VII & 3050 & 31 & 0.3321 & $19.35$ \\  \hline
Sphere & 2008 & 21 & 0.3321 & $9.53$\\ \hline
Aligned & 2008 & 21 & 0.3321 & 19.05 \\ \hline
\end{tabular}
\end{center}
\caption{
  Overview of the simulated parameter range: number of particles $N$,
  anisotropy $p$, packing fraction $\eta$, scaled droplet diameter $2
  R/L$.  Systems (I)-(VII) are planar, the system named ``sphere''
  corresponds to spherical geometry.  }
\label{TABparameters}
\end{table}

\end{document}